\documentclass[%
 preprint,
superscriptaddress,
 amsmath,amssymb,
 aps,prl
]{revtex4-2}
\usepackage[version=4]{mhchem}
\usepackage[colorlinks=true,linkcolor=black,citecolor=black,urlcolor=black]{hyperref}
\usepackage{bm}
\usepackage{graphicx}
\usepackage{amsmath}
\usepackage{multirow}
\usepackage{setspace}
\setcitestyle{numbers,square}
\makeatletter
\providecommand{\frontmatter@affiliationfont}{}
\renewcommand{\frontmatter@affiliationfont}{\small\itshape}
\renewcommand{\fnum@figure}{\textbf{\figurename~\thefigure}}
\makeatother

\begin{document}
\title{Unfolding Bloch States in Disordered Systems}

\author{T.~Thuy Hoang}
\affiliation{Department of Physics, Chungbuk National University, Cheongju 28644, Republic of Korea}
\affiliation{Advanced-Basic-Convergence Research Institute, Chungbuk National University, Cheongju 28644, Republic of Korea}

\author{Kunihiro Yananose}
\affiliation{Korea Institute for Advanced Study, Seoul 02455, Republic of Korea}

\author{Sungjong Woo}
\affiliation{Department of Physics, Ewha Womans University, Seoul 03760, Republic of Korea}
\affiliation{Institute for Multiscale Matter and Systems, Ewha Womans University, Seoul 03760, Republic of Korea}

\author{Seongjin Ahn}
\affiliation{Department of Physics, Chungbuk National University, Cheongju 28644, Republic of Korea}

\author{Dong Han}
\affiliation{College of Information Science and Engineering, Northeastern University, Shenyang 110819, People's Republic of China}

\author{Xian-Bin Li}
\thanks{Corresponding author: X.-B.~Li (lixianbin@jlu.edu.cn)}
\affiliation{State Key Laboratory of Integrated Optoelectronics, College of Electronic Science and Engineering, Jilin University, Changchun 130012, China}

\author{Junhyeok Bang}
\thanks{Corresponding author: J.~Bang (jbang@cbnu.ac.kr)}
\affiliation{Department of Physics, Chungbuk National University, Cheongju 28644, Republic of Korea}
\affiliation{Advanced-Basic-Convergence Research Institute, Chungbuk National University, Cheongju 28644, Republic of Korea}
\renewcommand{\abstractname}{}

\begin{abstract}
In crystalline solids, disorder breaks translational symmetry and obscures $\bm{k}$-resolved Bloch states,
limiting an accurate description of wavefunction-based observables. In this work, we present a method that
unfolds not only the band structures but also the corresponding Bloch states in disordered systems, going beyond
conventional band-unfolding techniques. As a prototype application, we study defective graphene and demonstrate
the capabilities by capturing key wavefunction-level responses, including disorder-driven redistribution
of Berry curvature.
\end{abstract}

\maketitle
\clearpage
\section*{Introduction}
Defects are ubiquitous and essentially unavoidable in real material \cite{Alkauskas2016}. Moreover, many technologically relevant systems—especially semiconductors—exhibit properties that are strongly altered by defects, while alloying provides a powerful route to systematically tune their electronic responses; therefore, reliable analysis of defected and alloyed materials is essential \cite{Zhu2007,Bang2009,Berlijn2011,Popescu2012,Rosenberger2018,Chen2016,Lee2014}. Such imperfections, however, break the translational symmetry of the pristine crystal and thereby complicate the $\bm{k}$-resolved band-structure picture, which remains one of the most informative and widely used representations of electronic states in crystalline solids \cite{Martin2020}. Nevertheless, a large body of experimental interpretation still relies on the phenomenological $\bm{k}$-resolved viewpoint, e.g., when extracting effective masses. In first-principles calculations, defects and alloys are typically treated using supercells (SCs) \cite{Freysoldt2014,Bang2016,Yazyev2010,Bang2010,Puska1998}. The use of an SC reduces the Brillouin zone (BZ) and folds the primitive-cell (PC) bands into the smaller SC BZ, leading to dense and overlapping dispersions. This obscures the underlying PC physics and makes direct comparison with band-structure-based experiments substantially more challenging \cite{Ku2010}.

Based on this need, band-unfolding methods were developed to restore an effective PC description from SC calculations \cite{Popescu2012,Popescu2010,Ku2010, Allen2013, Kim2008}. In widely used formulations, the unfolded band structure is obtained from a spectral function (spectral weight) constructed by projecting SC eigenstates onto PC Bloch states and collecting the resulting overlap weights into an effective $\bm{k}$-resolved spectral function in the PC BZ. This approach has enabled interpretation of folded SC bands and has been applied to a broad range of problems, including random alloys \cite{Popescu2012, Popescu2010}, defected/doped two-dimensional materials and adsorbed systems \cite{Medeiros2014}, and highly mismatched semiconductor alloys such as GaAsBi \cite{Maspero2017}. Despite this impact, conventional unfolding primarily returns energy dispersions and spectral weights in the PC BZ; it does not provide the corresponding unfolded Bloch eigenstates (i.e., wavefunctions with a well-defined $\bm{k}$-resolved structure). As a result, many physically important observables that are inherently wavefunction based—such as optical transition matrix elements (absorption strengths), orbital magnetic moments, and nonlinear optical coefficients (e.g., shift current) \cite{Gajdos2006,Hasan2010,Sipe2000}—cannot be straightforwardly accessed from the standard unfolding method. Especially, geometric and topological quantities including the Berry connection, Berry phase, and Berry curvature—whose far-reaching consequences in solid-state systems have attracted intense recent attention—require controlled $\bm{k}$-resolved Bloch wavefunctions \cite{Zak1989,Xiao2010}. Hence, they are not directly available in disordered systems, when unfolding provides only energies and spectral weights.

In this work, we develop a theoretical method that unfolds not only band dispersions and spectral weights, but also the corresponding Bloch eigenstates in disordered systems. Conventional unfolding schemes first diagonalize the SC Hamiltonian to obtain SC eigenstates and then project them onto PC Bloch states to construct spectral weights. Here we take the opposite route: we first project (represent) the SC Hamiltonian in the PC Bloch basis, yielding a $\bm{k}$-resolved decomposition of the disordered Hamiltonian. Diagonalizing each $\bm{k}$-block then directly produces the unfolded band structure together with well-defined unfolded Bloch states, while the inter-$\bm{k}$ blocks quantify disorder-induced mixing and thus determine the spectral broadening of the unfolded states. We demonstrate the method using disordered graphene, for which the $\pi$ Berry phase of Dirac electrons makes it a natural test platform for wavefunction-level geometric responses \cite{Hasan2010,zhang2005,Novoselov2005}. By considering symmetry-preserving and symmetry-breaking disorder, we capture the expected trends—such as bandgap opening and spreading of Berry curvature—while simultaneously predicting disorder-induced broadening within a unified description. Overall, our approach enables direct evaluation of wavefunction-based properties in disordered materials, offering a practical route to geometric and topological response calculations beyond standard band-unfolding outputs.

\section*{Theoretical Background}
\textbf{{\textit{k}-resolved decomposition of a disorder Hamiltonian}} -- For a disordered system, the total Hamiltonian $\hat{H}$ can be written as the sum of the pristine Hamiltonian $\hat{H}_0$ and a disorder contribution $\hat{H}_{\mathrm{dis}}$:
$\hat{H} = \hat{H}_0 + \hat{H}_{\mathrm{dis}}$.
Because $\hat{H}_0$ respects crystal translational symmetry, its eigenstates are Bloch states $| \bm{k}, n \rangle$ with well-defined crystal momentum:
$\hat{H}_0 | \bm{k}, n \rangle = \varepsilon_{\bm{k} n} | \bm{k}, n \rangle$. By introducing the projector onto the $\bm{k}$-subspace $\hat{P}_{\bm{k}} = \sum_n | \bm{k}, n \rangle \langle \bm{k}, n |$, we decompose $\hat{H}_{\mathrm{dis}}$  into a block-diagonal (intra-$\bm{k}$) part and an off-diagonal (inter-$\bm{k}$) part:
\begin{equation}
\hat{H}_{\mathrm{dis}}
= \sum_{\bm{k}} \hat{P}_{\bm{k}} \hat{H}_{\mathrm{dis}} \hat{P}_{\bm{k}}
+ \sum_{\bm{k} \neq\bm{k'}} \hat{P}_{\bm{k'}} \hat{H}_{\mathrm{dis}} \hat{P}_{\bm{k}}
\equiv
\hat{H}_{\mathrm{dis}}^{\bm{k}}
+ \hat{H}_{\mathrm{dis}}^{\bm{k'},\bm{k}} .
\label{1}
\end{equation}
Here, $\hat{H}_{\mathrm{dis}}^{\bm{k}}$ is block-diagonal within each $\bm{k}$, 
whereas $\hat{H}_{\mathrm{dis}}^{\bm{k'},\bm{k}}$ couples different crystal momenta 
$\bm{k} \neq\bm{k'}$. 
This $\bm{k}$-resolved decomposition allows us to obtain unfolded Bloch states and their spectral weights on a consistent footing.

We argue that $\bigl\| \hat{H}_{\mathrm{dis}}^{\bm{k'},\bm{k}} \bigr\| \ll\bigl\| \hat{H}_{\mathrm{dis}}^{\bm{k}} \bigr\|$, and, in fact, $\hat{H}_{\mathrm{dis}}^{\bm{k'},\bm{k}} \to 0$ under randomly distributed disorder, as fully discussed in the Supplementary Material (SM) \cite{SupplementalMaterial}. Assuming multiple defect types and mutually independent defects, we write  
$
H_{\mathrm{dis}}(\bm{r})
=
\sum_d
\left[
\sum_{i_d=1}^{N_d}
V_d(\bm{r} - \bm{R}_{i_d})
\right]
$, where $N_d$ is the number of defects of type $d$, 
$\bm{R}_{i_d}$ is the lattice vector of the unit cell containing the $i$-th $d$-type defect, 
and $V_d$ is the corresponding single-defect potential. With this form, the intra-$\bm{k}$ matrix elements
$
\langle \bm{k}, n | \hat{H}_{\mathrm{dis}} | \bm{k}, m \rangle
$
add \textit{coherently} over defect positions and therefore remain finite scaling with the defect concentration $\rho_d = N_d / N$, where $N$ is the number of unit cells. 
In contrast, the inter-$\bm{k}$ matrix elements 
$\langle \bm{k'}, n | \hat{H}_{\mathrm{dis}} | \bm{k}, m \rangle$ 
($\bm{k'} \neq\bm{k}$) acquire an additional phase factor 
$e^{i(\bm{k}-\bm{k'}) \cdot \bm{R}_{i_d}}$ from each defect position. 
For randomly distributed defects, these phases are \textit{uncorrelated and largely cancel} in the sum over $\bm{R}_{i_d}$, so that the inter-$\bm{k}$ block vanishes under ensemble averaging. 
Consequently, the disorder Hamiltonian is dominated by its $\bm{k}$-diagonal blocks, while the $\bm{k'} \neq\bm{k}$ blocks are parametrically small in the thermodynamic limit. In light of the above, we decompose the full Hamiltonian,
$
\hat{H} = \hat{H}_0 + \hat{H}_{\mathrm{dis}}^{\bm{k}} + \hat{H}_{\mathrm{dis}}^{\bm{k'},\bm{k}}
$,
and regard $\hat{H}_0 + \hat{H}_{\mathrm{dis}}^{\bm{k}}$ as the dominant part while treating 
$\hat{H}_{\mathrm{dis}}^{\bm{k'},\bm{k}}$ as a perturbation. 
The dominant block defines unfolded states, i.e., \textit{dressed Bloch states}, and the off-diagonal term 
$\hat{H}_{\mathrm{dis}}^{\bm{k'},\bm{k}}$ induces inter-$\bm{k}$ mixing between these states, thereby generating a finite spectral weight in the unfolded representation.

Although this work considers only the diagonal components of the disorder Hamiltonian for the renormalization of the dressed Bloch states, the off-diagonal components $\hat{H}_{\mathrm{dis}}^{\bm{k'},\bm{k}}$ would provide further high order corrections, mixing the Bloch states at different momenta  \cite{Li2009,Groth2009,Chen2022,Liu2024}. 
Such higher-order effects can be systematically investigated within the framework of many-body perturbation theory, where one treats $\hat{H}_0 + \hat{H}_{\mathrm{dis}}^{\bm{k}}$ as the unperturbed bare Hamiltonian and determines the interacting Green’s functions by solving Dyson’s equation \cite{Mahan2013}. 
While we can further refine the validity of our theory by incorporating higher-order Feynman diagrams into the evaluation of the self-energy, we expect such corrections are quantitatively marginal because the off-diagonal perturbing term  $\hat{H}_{\mathrm{dis}}^{\bm{k'},\bm{k}}$ is sufficiently weak relative to the unperturbed Hamiltonian $(\hat{H}_0 + \hat{H}_{\mathrm{dis}}^{\bm{k}})$, as discussed in the previous paragraph. 
Thus, in the following we retain only the dominant leading diagonal term  $\hat{H}_{\mathrm{dis}}^{\bm{k}}$ for the electronic renormalization.

\textbf{{Dressed Bloch states in a disordered system}} -- In the leading term $\hat{H}_0 + \hat{H}_{\mathrm{dis}}^{\bm{k}}$, 
$\hat{H}_0$ is already diagonal in the Bloch basis 
$| \bm{k}, n \rangle$. 
While $\hat{H}_{\mathrm{dis}}^{\bm{k}}$ is not diagonal, 
it is block-diagonal in $\bm{k}$. 
Consequently, we can diagonalize each $\bm{k}$ block independently 
to obtain the dressed Bloch states 
$| \widetilde{\bm{k}, n} \rangle$ 
with the same good quantum number $\bm{k}$:
\begin{equation}
\left( \hat{H}_0 + \hat{H}_{\mathrm{dis}}^{\bm{k}} \right)
| \widetilde{\bm{k}, n} \rangle
=
\tilde{\varepsilon}_{\bm{k}, n}
| \widetilde{\bm{k}, n} \rangle ,
\label{2}
\end{equation}
where $\tilde{\varepsilon}_{\bm{k}, n}$ is the dressed energy eigenvalue 
for $| \widetilde{\bm{k}, n} \rangle$. 
This procedure avoids diagonalizing a single large matrix, 
which is typically required in conventional band-unfolding approaches \cite{Popescu2012,Ku2010,Popescu2010}. 
Instead, our method requires only diagonalizing small matrices for each $\bm{k}$, 
yielding substantial computational savings. 
Physically, $| \widetilde{\bm{k}, n} \rangle$ is a long-lived stationary state 
in the weak-disorder limit, and more generally it serves as a quasi-eigenstate 
on intermediate timescales between scattering events. 
As such, physical properties driven by the dressed Bloch state 
are valid within the timescale.

\textbf{Estimating the spectral weight by disorders} -- Each individual defect acts as a scatterer, 
and the per-$d$ defect transition rate 
$\Gamma^{d}_{\bm{k},n \to \bm{k'},m}$ 
from an initial dressed state 
$| \widetilde{\bm{k}, n} \rangle$ 
to a final dressed state 
$| \widetilde{\bm{k'}, m} \rangle$ 
is given by Fermi's golden rule \cite{Sakurai2020},
\begin{equation}
\Gamma^{d}_{\bm{k},n \to \bm{k'},m}
=
\frac{2\pi}{\hbar}
\left|
\langle \widetilde{\bm{k'}, m} |
\hat{V}_d
| \widetilde{\bm{k}, n} \rangle
\right|^2
\delta \left(
\tilde{\varepsilon}_{\bm{k'},m}
-
\tilde{\varepsilon}_{\bm{k},n}
\right),
\label{3}
\end{equation}
where the delta function enforces elastic scattering. 
The total rate $\bar{\Gamma}_{\bm{k},n}$ at which a dressed state 
$| \widetilde{\bm{k}, n} \rangle$ 
scatters into \textit{all} final states due to \textit{all} defects is
\begin{equation}
\bar{\Gamma}_{\bm{k},n}
=
\sum_{d,\bm{k'},m}
N_d \,
\Gamma^{d}_{\bm{k},n \to \bm{k'},m}.
\label{4}
\end{equation}
The detailed formula is discussed in the SM \cite{SupplementalMaterial}. 
Similar formulas have appeared in prior work \cite{Lordi2010,Kim2019}, 
but here we evaluate them using the \textit{dressed} Bloch states 
$| \widetilde{\bm{k}, n} \rangle$ 
rather than the pristine Bloch states $| \bm{k}, n \rangle$. $\bar{\Gamma}_{\bm{k},n}$ is physically related to the scattering time 
$\tau_{\bm{k},n} = \frac{1}{\bar{\Gamma}_{\bm{k},n}}$, 
so it can be interpreted as the spectral weight, i.e., the broadening, of the 
$| \widetilde{\bm{k}, n} \rangle$ state.

\section*{Application to Disordered Graphene}

\begin{figure}[htbp]
    \centering
    \includegraphics[width=0.9\linewidth]{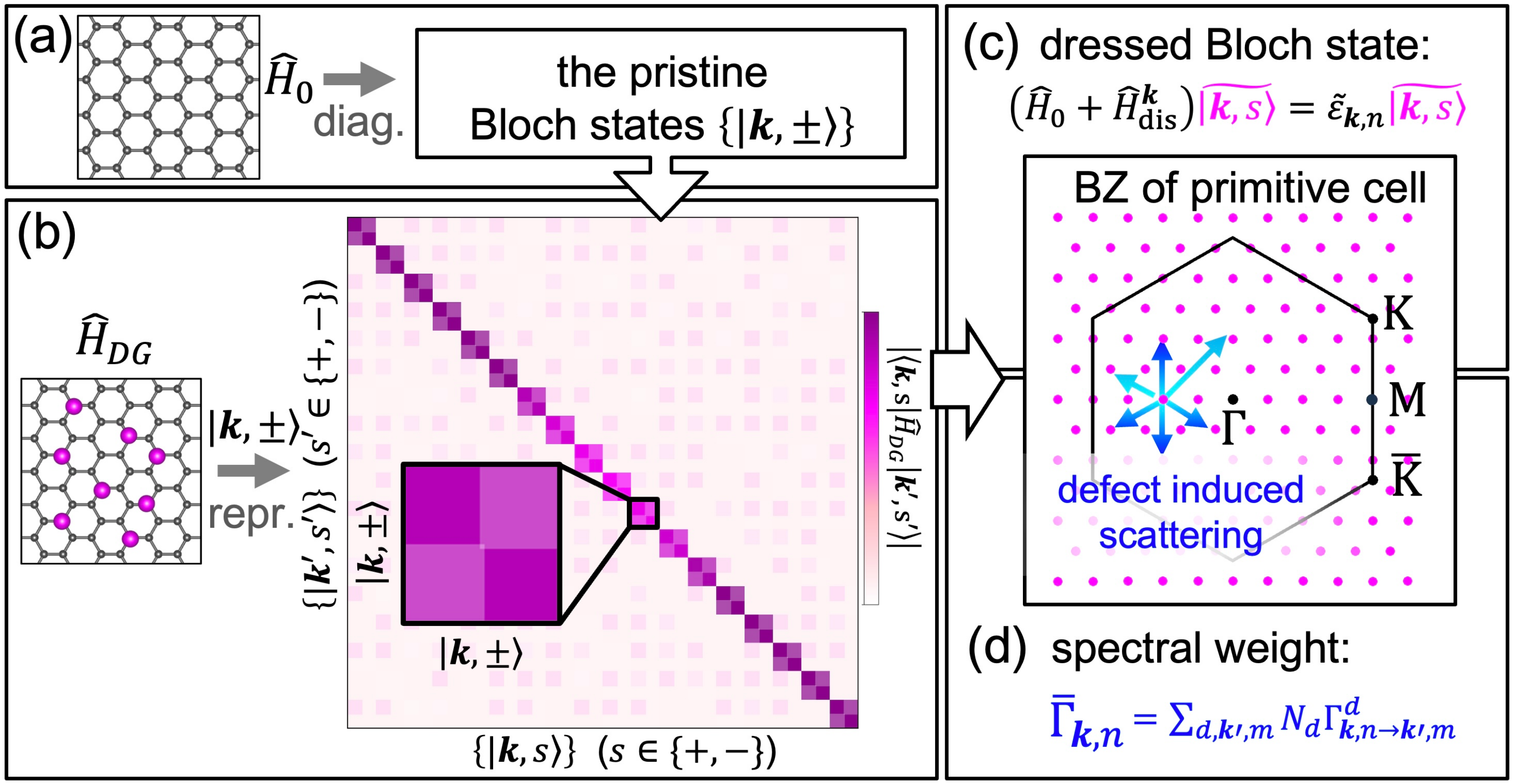}
    \caption{Workflow for unfolding Bloch states in disordered graphene.
(a) Diagonalization of the pristine Hamiltonian $\hat{H}_0$ at each $\bm{k}$-point
in the primitive-cell (PC) Brillouin zone (BZ), yielding the pristine Bloch
eigenstate $|\bm{k},\pm\rangle$.
(b) Representation of the defect Hamiltonian $\hat{H}_{DG}$ using
$|\bm{k},\pm\rangle$, i.e.,
$\langle \bm{k},s|\hat{H}_{DG}|\bm{k'},s'\rangle$ with $s,s' \in \{+,-\}$.
(c) Block-diagonalization of $\hat{H}_{DG}$ for a given $\bm{k}$ to find the
dressed Bloch energies $\tilde{\varepsilon}_{\bm{k},n}$ and dressed Bloch states
$|\widetilde{\bm{k},\pm}\rangle$.
(d) Evaluating the spectral weight $\Gamma_{\bm{k},n}$ using the calculated
dressed Bloch states $|\widetilde{\bm{k},\pm}\rangle$.
}
    \label{fig1}
\end{figure}

\begin{figure}[htbp]
    \centering
    \includegraphics[width=0.6\linewidth]{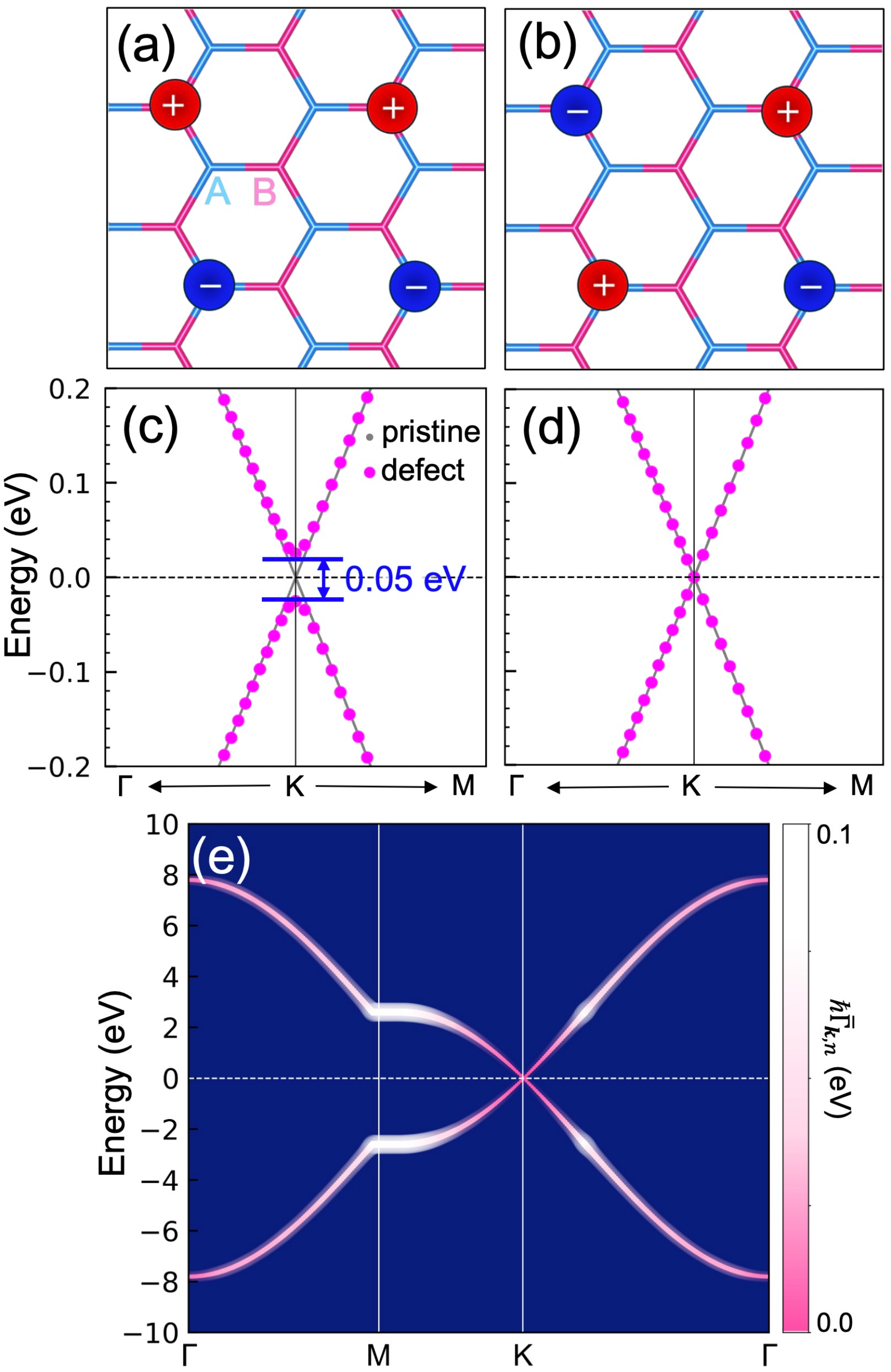}
    \caption{Disordered-graphene models and their unfolded electronic structures. (a) Symmetry-breaking (SB) disorder, where positive (red) and negative (blue) on-site energies are assigned exclusively to the A and B sublattices, respectively. (b) Symmetry-preserving (SP) disorder, where each sublattice contains equal numbers of the positive and negative on-site energy. (c,d) Dressed band structure $\tilde{\varepsilon}_{\bm{k}, n}$ for (c) SB and (d) SP disorder (magenta), shown together with the pristine graphene bands (gray) for comparison. (e) Unfolded band structure for the SP case including spectral broadening ($\rho = 5\%$).
}
    \label{fig2}
\end{figure}

As shown in Fig.~\ref{fig1}, we apply the above framework to disordered graphene as a demonstrative example. 
A graphene sheet with random disorder is modeled within a nearest-neighbor tight-binding Hamiltonian,
\begin{equation}
\hat{H}_{\mathrm{DG}}
=
t \sum_{\langle l m \rangle} 
\hat{C}_l^\dagger \hat{C}_m
+
\sum_n U_n \hat{C}_n^\dagger \hat{C}_n ,
\label{5}
\end{equation}
where $t$ ($= \text{–}2.6~\mathrm{eV}$) is the nearest-neighbor hopping energy between carbon $\pi$-orbitals, 
$\hat{C}_l$ ($\hat{C}_l^\dagger$) annihilates (creates) an electron on site $l$, 
and $U_n$ denotes the on-site disorder energy at site $n$. 
In Eq.~(\ref{5}), the first term is the pristine graphene Hamiltonian ($\hat{H}_0$), and the second term defines the disorder potential ($\hat{H}_{\mathrm{dis}}$). The simulation uses a $100 \times 100$ SC of the primitive honeycomb lattice 
($2 \times 10^4$ carbon sites), which ensures the randomness of defects. Defects are introduced by randomly selecting carbon sites at a concentration of $\rho = 1\%$ and $5\%$. 
The on-site potential is set to $U_n = 0$~eV on pristine sites, 
while defect sites are assigned $U_n = \text{+}0.5$~eV or $\text{–}0.5$~eV with equal probability. We consider two disorder realizations, illustrated in Figs.~\ref{fig2}(a) and \ref{fig2}(b). In the symmetry-breaking (SB) disorder [Fig. \ref{fig2}(a)], the positive ($\text{+}0.5$~eV) and negative ($\text{–}0.5$~eV) 
potentials are placed exclusively on the A and B sublattices, respectively, thereby inducing sublattice asymmetry. In the symmetry-preserving (SP) disorder [Fig. \ref{fig2}(b)], each sublattice contains equal numbers of $\text{+}0.5$~eV and $\text{–}0.5$~eV sites, so that the average on-site potential vanishes separately on both the A and B sublattices. 

Our computational workflow, illustrated in Fig. \ref{fig1}, consists of four steps: \textit{(1) Diagonalize the pristine Hamiltonian $\hat{H}_0$} [Fig. \ref{fig1}(a)].  
For each $\bm{k}$ in the PC BZ, we solve
$
\hat{H}_0 | \bm{k}, \pm \rangle
=
\varepsilon_{\bm{k},\pm}
| \bm{k}, \pm \rangle ,
$
where $\text{+}$ and $-$ label the conduction and valence bands of graphene, respectively.  This step requires only the PC and provides the band structure of pristine graphene, as shown in Figs. \ref{fig2}(c) and \ref{fig2}(d).

\medskip

\textit{(2) Represent the total Hamiltonian $\hat{H}_{\mathrm{DG}}$ using the pristine Bloch basis} [Fig.~\ref{fig1}(b)].  
To model defective graphene, we employ a SC and represent $\hat{H}_{\mathrm{DG}}$ in the $| \bm{k}, \pm \rangle$ basis, i.e.,
$
\langle \bm{k}, s | \hat{H}_{\mathrm{DG}} | \bm{k'}, s' \rangle$, with $ s,s' \in \{+,-\}
$. In our calculations, the resulting matrix is dominated by intra-$\bm{k}$ blocks
($
\langle \bm{k}, s | \hat{H}_{\mathrm{DG}} | \bm{k}, s' \rangle
$),
and has numerically negligible inter-$\bm{k}$ elements (mostly below $5.0 \times 10^{-4}$ eV) [see Fig. \ref{fig2}(b)].  
This is consistent with the discussion above, i.e.,
$
\left\| \hat{H}_{\mathrm{dis}}^{\bm{k'},\bm{k}} \right\|
\ll
\left\| \hat{H}_{\mathrm{dis}}^{\bm{k}} \right\|
$.
Accordingly, the inter-$\bm{k}$ terms can be treated within perturbation theory.
\medskip

\textit{(3) Block-diagonalize $\hat{H}_{\mathrm{DG}}$ for a given $\bm{k}$} [Fig. \ref{fig1}(c)]. For each $\bm{k}$, we diagonalize the corresponding small block (here $2\times2$; in general $n\times n$ with $n$ bands retained) to obtain the dressed Bloch energies $\tilde{\varepsilon}_{\bm{k},n}$ and states $| \widetilde{\bm{k}, \pm} \rangle$, following Eq.~(\ref{2}). Figures~\ref{fig2}(c) and \ref{fig2}(d) show the renormalized band structures (dressed eigenvalues $\tilde{\varepsilon}_{\bm{k},n}$) for SB and SP disordered graphene, respectively.  
In the SB case, disorder induces a band gap of about $0.05$ eV [Fig.~\ref{fig2}(c)].  
The resulting band structure closely resembles that of graphene with a uniform sublattice-staggered on-site potential (a mass term) of $\pm 0.025$ eV, consistent with the disorder-averaged scale.  
In contrast, the SP disorder remains gapless and closely follows the pristine dispersion [Fig.~\ref{fig2}(d)], in good agreement with previous work \cite{Kang2008}.

\begin{figure}[htbp]
    \centering
    \includegraphics[width=0.9\linewidth]{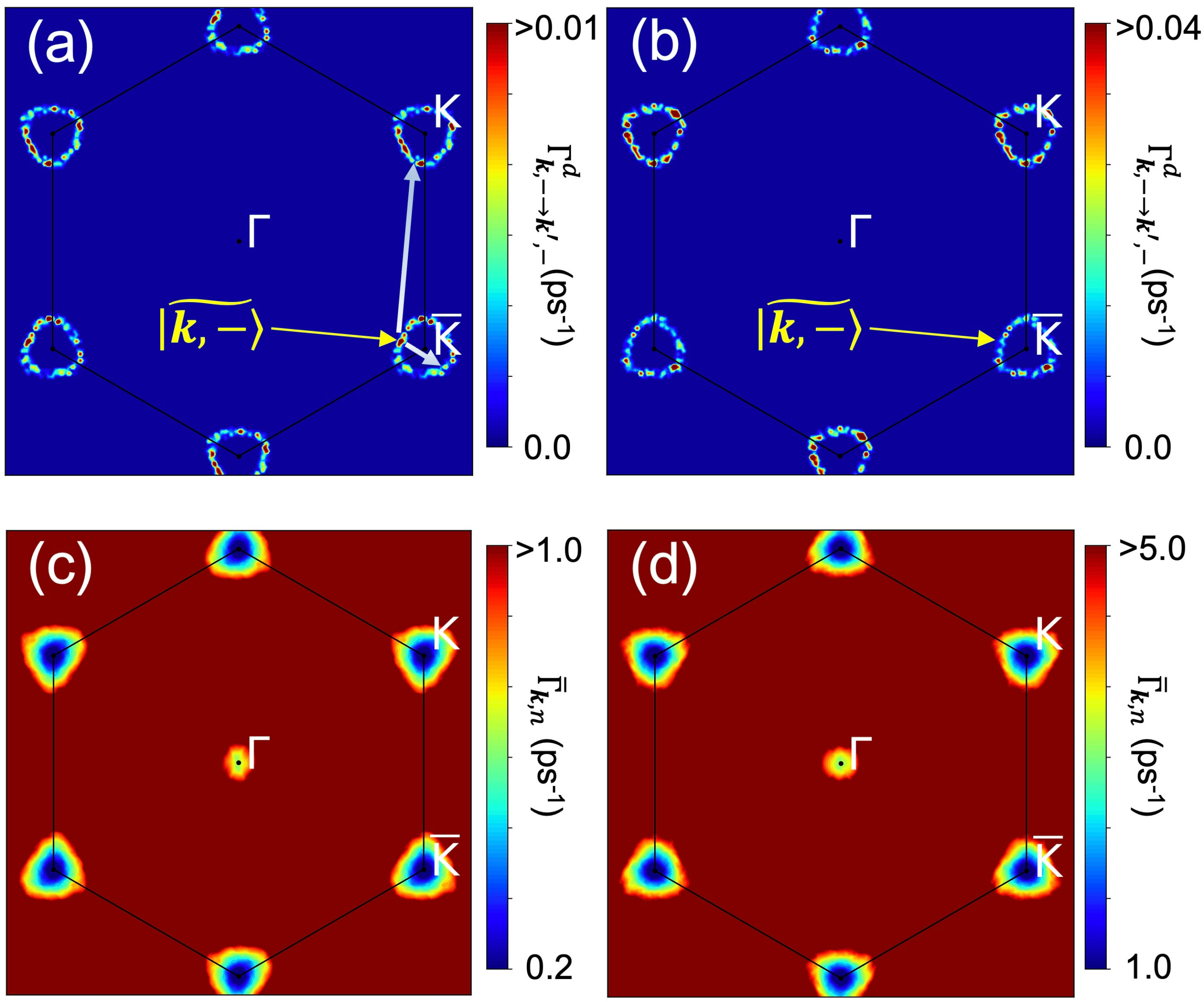}
    \caption{Disorder-induced scattering and spectral broadening in disordered graphene (SP disorder).
Momentum-resolved scattering rate $\Gamma^{d}_{\bm{k},-\rightarrow \bm{k'},-}$ from a fixed initial dressed state $|\widetilde{\bm{k},-}\rangle$ chosen near the $\overline{K}$ valley (indicated by the yellow arrow) for defect concentrations (a) $\rho = 1\%$ and (b) $\rho = 5\%$. The arrows in (a) highlight representative intra- and inter-valley scatterings. Distribution of the total spectral broadening $\Gamma_{\bm{k} ,n}$ over the first Brillouin zone for (c) $\rho = 1\%$ and (d) $\rho = 5\%$.
}
    \label{fig3}
\end{figure}

\textit{(4) Spectral weight (broadening) of the dressed Bloch state $| \widetilde{\bm{k}, \pm} \rangle$} [Fig.~\ref{fig1}(d)].  
Using Eqs.~(\ref{3}) and (\ref{4}), we calculate $\bar{\Gamma}_{\bm{k},n}$ for $| \widetilde{\bm{k}, n} \rangle$, which is considered as the spectral weight of the corresponding state. Figures~ \ref{fig3}(a) and \ref{fig3}(b) show the scattering rate 
$\Gamma^{d}_{\bm{k},- \rightarrow \bm{k'},-}$ 
from a given initial state $|\bm{k},-\rangle$ near the $\overline{K}$ valley 
[marked in Figs.~\ref{fig3}(a) and \ref{fig3}(b)] for $\rho = 1\%$ and $5\%$, respectively. 
$\Gamma^{d}_{\bm{k},- \rightarrow \bm{k'},-}$ is very similar for the SB and SP cases; 
therefore, we present only the SP results in what follows. 
The results show that disorder induces both intravalley and intervalley scatterings, 
and because the scattering is elastic, nonzero 
$\Gamma^{d}_{\bm{k},- \rightarrow \bm{k'},-}$ 
appears as ring-like features around the $K$ and $\overline{K}$ valleys. 
The typical value of 
$\Gamma^{d}_{\bm{k},- \rightarrow \bm{k'},-}$ 
for $\rho = 1\%$ is about $0.004\,\text{ps}^{-1}$, 
which is approximately five times smaller than that for $\rho = 5\%$ 
($\sim 0.02\,\text{ps}^{-1}$), consistent with the scaling with defect density. By summing over all final states in 
$\Gamma^{d}_{\bm{k},- \rightarrow \bm{k'},-}$ [Eq.~(\ref{4})], 
we obtain $\overline{\Gamma}_{\bm{k},n}$, i.e., the total broadening of 
$|\bm{k},\pm\rangle$, as shown in Fig. \ref{fig3}(c) for $\rho = 1\%$ and \ref{fig3}(d) for $\rho = 5\%$. 
Similar to $\Gamma^{d}_{\bm{k},- \rightarrow \bm{k'},-}$, 
$\overline{\Gamma}_{\bm{k},n}$ increases with $\rho$ 
(see the color-bar ranges in Fig. \ref{fig3}). 
In both cases, $\overline{\Gamma}_{\bm{k},n}$ is small near the $K$ and 
$\overline{K}$ points. In particular, it remains below 
0.2 $\text{ps}^{-1}$ at $\rho = 1\%$ and 
1 $\text{ps}^{-1}$ at $\rho = 5\%$ in the vicinity of the valleys.
Using $\overline{\Gamma}_{\bm{k},n}$, we plot the unfolded band structure 
for the SP disorder including spectral broadening, as shown in Fig. \ref{fig2}(e), 
where the broadening is largest near the \textit{M} point, consistent with the higher density of states.

\begin{figure}[htbp]
    \centering
    \includegraphics[width=0.9\linewidth]{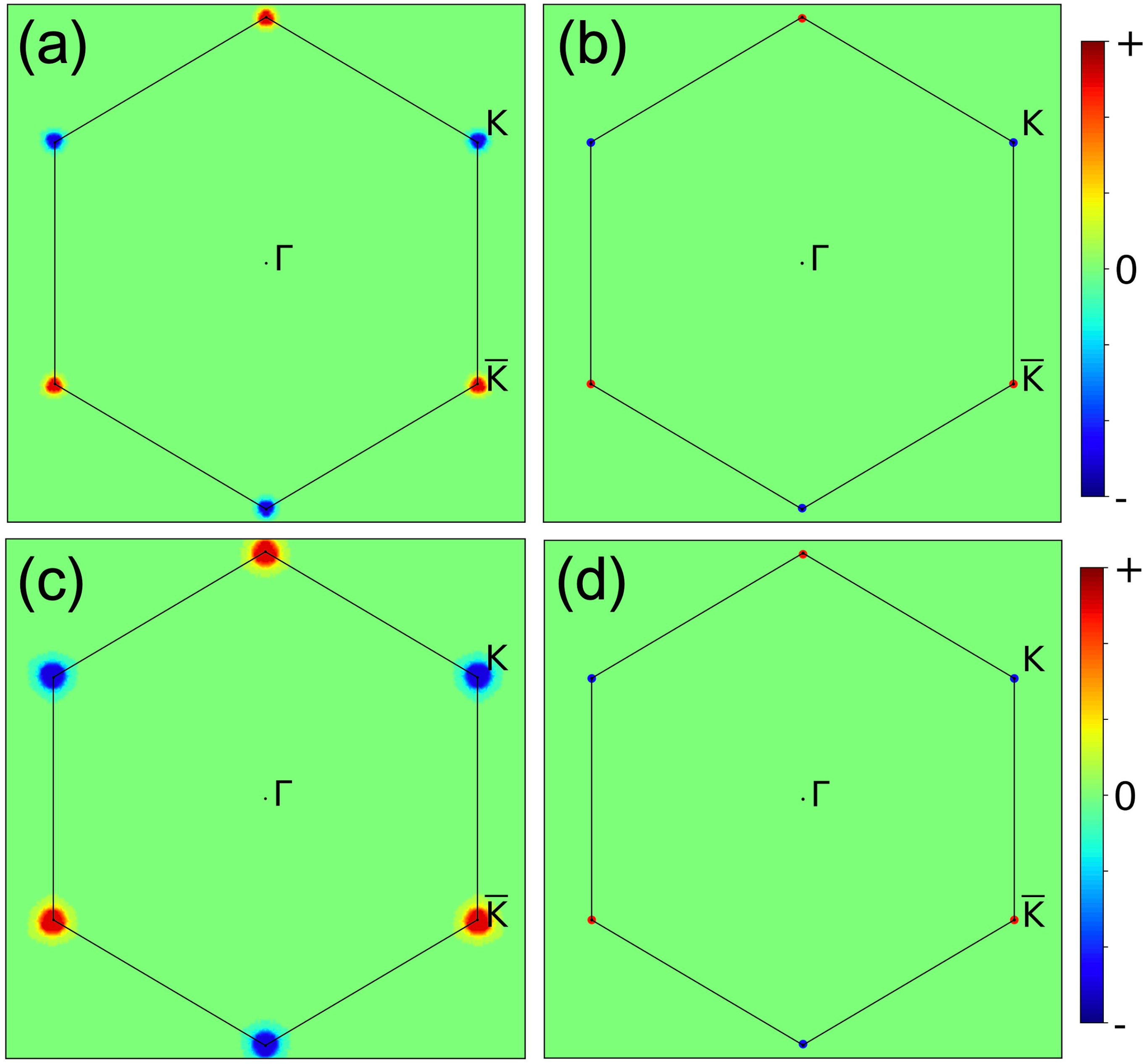}
    \caption{Distribution of Berry curvature in the Brillouin zone for the SB and SP disorders.
Panels (a,c) show SB disorder and panels (b,d) show SP disorder at defect concentrations (a,b) $\rho = 1\%$ and (c,d) $\rho = 5\%$.
The color scale indicates the sign and magnitude of the Berry curvature, with contributions concentrated near the valley points \textit{K} and $\overline{K}$.
}
    \label{fig4}
\end{figure}

Because we obtain the unfolded Bloch states $|\widetilde{\bm{k}, \pm} \rangle$, 
we can directly evaluate wavefunction-based observables. 
Here we focus on the Berry curvature (BC) as a representative example of such quantities \cite{Resta1994}. 
Since $| \widetilde{\bm{k}, \pm} \rangle$ is well defined in the primitive-cell $\bm{k}$-space, 
we can employ the standard BC formalism developed for crystalline systems \cite{Xiao2010, Resta1994,Resta2010,Tyner2024,PythTB2022,Vanderbilt2006}. The numerical procedure is detailed in SM \cite{SupplementalMaterial}. 
Figure \ref{fig4} shows the resulting BC distribution. 
As expected, nonzero BC is concentrated around the $K$ and $\overline{K}$ valleys  
in both SB and SP cases. As in pristine graphene and related two-dimensional materials 
(e.g., transition-metal dichalcogenides), BC has opposite signs at the $K$ and 
$\overline{K}$ valleys. 
In the SB case, BC around the valleys is no longer sharply localized and becomes spatially broadened [Figs. \ref{fig4}(a) and \ref{fig4}(c)], and the broadening increases with defect concentration.
In contrast, in the SP case BC remains well localized near the valleys, 
consistent with the robustness of the nontrivial $\pi$ Berry phase 
[Figs. \ref{fig4}(b) and \ref{fig4}(d)]. These contrasting behaviors are consistent with the presence  (SB) or absence (SP) of a band gap, reflecting whether inversion symmetry is effectively broken \cite{Xiao2007,Yao2008}. We note that earlier first-principles studies combined BC calculations with band unfolding by computing BC in a defective SC treated as an artificially periodic crystal and then unfolding the SC BZ onto the PC BZ \cite{Bianco2014,Olsen2015,Martiny2019}. Because BC is evaluated in the SC $\bm{k}$-space and only mapped afterward, the procedure is not a direct calculation in the PC $\bm{k}$-parameter space and is no longer geometric, i.e., the usual interpretation as a Berry phase per unit area is lost. In contrast, our method constructs unfolded Bloch states 
defined on the PC BZ, enabling BC to be computed directly in the proper $\bm{k}$-space where geometric quantities are formulated.

\section*{Summary}
In summary, we have presented a theoretical framework that produces both the unfolded band structure and the corresponding unfolded Bloch states for disordered systems, overcoming the limitations of previous approaches, which yield only energies and spectral weights. Our method proceeds in two key steps: (1) representing the SC Hamiltonian onto the PC Bloch basis and (2) diagonalizing the resulting $\bm{k}$-resolved blocks. By reversing the conventional unfolding workflow, this construction directly delivers unfolded Bloch eigenstates, with their spectral broadening governed by the inter-$\bm{k}$ couplings. Consequently, our approach enables the direct evaluation of wavefunction-level observables in disordered materials and provides a practical route to geometric and topological response calculations in the proper primitive-cell $\bm{k}$-space, beyond the scope of conventional band-unfolding methods.

\section*{ACKNOWLEDGMENT}
This work was supported by Basic Science Research Program through the National Research Foundation of Korea (NRF) (RS-2023-NR076774), Global-Learning \& Academic research institution for Master’s·PhD students, and Postdocs (LAMP) program (RS-2024-00445180), and Information Technology Research Center (IITP-RS-2024-00437284). This work was also supported by Chungbuk National University NUDP program (2024). D. H. acknowledges the support by National Natural Science Foundation of China (Grants No. 12474079), Scientific and Technological Development Project of Jilin Province, China (Grant No. 20230101004JC), and the Open Fund of the State Key Laboratory of Luminescence Science and Technology (Grant No. SKLA-2024-01). X. L. acknowledges the support by National Natural Science Foundation of China (Grant No. 12274172). K.Y. was supported by a grant at Korea Institute for Advanced Study (CG092502).

\clearpage

\clearpage
\clearpage
\clearpage

 \clearpage    
 
\bibliography{apssamp}

@article{Alkauskas2016,
  author  = {Alkauskas, A. and McCluskey, M. D. and {Van de Walle}, C. G.},
  title   = {Tutorial: Defects in semiconductors---Combining experiment and theory},
  journal = {J. Appl. Phys.},
  volume  = {119},
  pages   = {181101},
  year    = {2016},
  doi     = {10.1063/1.4948245}
}

@article{Zhu2007,
  title = {Nanoscale Disorder in \ce{${\mathrm{CaCu}}_{3}{\mathrm{Ti}}_{4}{\mathrm{O}}_{12}$}: A New Route to the Enhanced Dielectric Response},
  author = {Zhu, Y. and Zheng, J. C. and Wu, L. and Frenkel, A. I. and Hanson, J. and Northrup, P. and Ku, W.},
  journal = {Phys. Rev. Lett.},
  volume = {99},
  issue = {3},
  pages = {037602},
  numpages = {4},
  year = {2007},
  month = {Jul},
  publisher = {American Physical Society},
  doi = {10.1103/PhysRevLett.99.037602},
  url = {https://link.aps.org/doi/10.1103/PhysRevLett.99.037602}
}

@article{Bang2009,
   author = {Bang, Junhyeok and Choi, Eun-Ae and Chang, K. J.},
    title = {The effect of impurities on hydrogen bonding site and local vibrational frequency in \ce{ZnO}},
    journal = {J. Appl. Phys.},
    volume = {106},
    number = {5},
    pages = {053522},
    year = {2009},
    month = {09},
    issn = {0021-8979},
    doi = {10.1063/1.3213387},
    url = {https://doi.org/10.1063/1.3213387},
}

@article{Berlijn2011,
  title = {Can Disorder Alone Destroy the ${e}_{g}^{\ensuremath{'}}$ Hole Pockets of \ce{${\mathrm{Na}}_{x}{\mathrm{CoO}}_{2}$}? A \ce{W}annier Function Based First-Principles Method for Disordered Systems},
  author = {Berlijn, Tom and Volja, Dmitri and Ku, Wei},
  journal = {Phys. Rev. Lett.},
  volume = {106},
  issue = {7},
  pages = {077005},
  numpages = {4},
  year = {2011},
  month = {Feb},
  publisher = {American Physical Society},
  doi = {10.1103/PhysRevLett.106.077005},
  url = {https://link.aps.org/doi/10.1103/PhysRevLett.106.077005}
}

@article{Popescu2012,
  title = {Extracting \ce{$E$} versus $\vec{k}$  effective band structure from supercell calculations on alloys and impurities},
  author = {Popescu, Voicu and Zunger, Alex},
  journal = {Phys. Rev. B},
  volume = {85},
  issue = {8},
  pages = {085201},
  numpages = {12},
  year = {2012},
  month = {Feb},
  publisher = {American Physical Society},
  doi = {10.1103/PhysRevB.85.085201},
  url = {https://link.aps.org/doi/10.1103/PhysRevB.85.085201}
}

@article{Rosenberger2018,
author = {Rosenberger, Matthew R. and Chuang, Hsun-Jen and McCreary, Kathleen M. and Li, Connie H. and Jonker, Berend T.},
title = {Electrical Characterization of Discrete Defects and Impact of Defect Density on Photoluminescence in Monolayer \ce{$\mathrm{WS_2}$}},
journal = {ACS Nano},
volume = {12},
number = {2},
pages = {1793-1800},
year = {2018},
doi = {10.1021/acsnano.7b08566},
URL = { 
        https://doi.org/10.1021/acsnano.7b08566
}
  
}

@article{Chen2016,
  author = {Chen, M. and Ge, Z. and Li, Y. and Agterberg, D. and Li, L. and Weinert, M.},
  title = {Effects of interface oxygen vacancies on electronic bands of \ce{Fe/SrTiO3 (001})},
  journal = {Phys. Rev. B},
  volume = {94},
  pages = {245139},
  year = {2016},
  publisher = {American Physical Society},
  doi = {10.1103/PhysRevB.94.245139},
  url = {https://link.aps.org/doi/10.1103/PhysRevB.94.245139}
  
}

@article{Lee2014,
  title = {Competing magnetism in $\ensuremath{\pi}$-electrons in graphene with a single carbon vacancy},
  author = {Lee, Chi-Cheng and Yamada-Takamura, Yukiko and Ozaki, Taisuke},
  journal = {Phys. Rev. B},
  volume = {90},
  issue = {1},
  pages = {014401},
  numpages = {5},
  year = {2014},
  month = {Jul},
  publisher = {American Physical Society},
  doi = {10.1103/PhysRevB.90.014401},
  url = {https://link.aps.org/doi/10.1103/PhysRevB.90.014401}
}

@book{Martin2020,
  author = {Martin, R. M.},
  title = {Electronic Structure: Basic Theory and Practical Methods},
  publisher = {Cambridge University Press},
  year = {2020}
}

@article{Freysoldt2014,
  title = {First-principles calculations for point defects in solids},
  author = {Freysoldt, Christoph and Grabowski, Blazej and Hickel, Tilmann and Neugebauer, J\"org and Kresse, Georg and Janotti, Anderson and Van de Walle, Chris G.},
  journal = {Rev. Mod. Phys.},
  volume = {86},
  issue = {1},
  pages = {253--305},
  numpages = {53},
  year = {2014},
  month = {Mar},
  publisher = {American Physical Society},
  doi = {10.1103/RevModPhys.86.253},
  url = {https://link.aps.org/doi/10.1103/RevModPhys.86.253}
}

@article{Bang2016,
  author = {Bang, J. and Sun, Y. Y. and Song, J.-H. and Zhang, S. B.},
  title = {Carrier-induced transient defect mechanism for non-radiative recombination in \ce{InGaN} light-emitting devices},
  journal = {Sci. Rep.},
  volume = {6},
  pages = {24404},
  year = {2016},
  doi = {10.1038/srep24404},
  url = {https://www.nature.com/articles/srep24404#citeas}
}

@article{Yazyev2010,
 title = {Topological defects in graphene: \ce{D}islocations and grain boundaries},
  author = {Yazyev, Oleg V. and Louie, Steven G.},
  journal = {Phys. Rev. B},
  volume = {81},
  issue = {19},
  pages = {195420},
  numpages = {7},
  year = {2010},
  month = {May},
  publisher = {American Physical Society},
  doi = {10.1103/PhysRevB.81.195420},
  url = {https://link.aps.org/doi/10.1103/PhysRevB.81.195420}
}

@article{Bang2010,
title = {Localization and one-parameter scaling in hydrogenated graphene},
  author = {Bang, Junhyeok and Chang, K. J.},
  journal = {Phys. Rev. B},
  volume = {81},
  issue = {19},
  pages = {193412},
  numpages = {4},
  year = {2010},
  month = {May},
  publisher = {American Physical Society},
  doi = {10.1103/PhysRevB.81.193412},
  url = {https://link.aps.org/doi/10.1103/PhysRevB.81.193412}
}

@article{Puska1998,
  title = {Convergence of supercell calculations for point defects in semiconductors: Vacancy in silicon},
  author = {Puska, M. J. and P\"oykk\"o, S. and Pesola, M. and Nieminen, R. M.},
  journal = {Phys. Rev. B},
  volume = {58},
  issue = {3},
  pages = {1318--1325},
  numpages = {0},
  year = {1998},
  month = {Jul},
  publisher = {American Physical Society},
  doi = {10.1103/PhysRevB.58.1318},
  url = {https://link.aps.org/doi/10.1103/PhysRevB.58.1318}
}

@article{Ku2010,
  title = {Unfolding First-Principles Band Structures},
  author = {Ku, Wei and Berlijn, Tom and Lee, Chi-Cheng},
  journal = {Phys. Rev. Lett.},
  volume = {104},
  issue = {21},
  pages = {216401},
  numpages = {4},
  year = {2010},
  month = {May},
  publisher = {American Physical Society},
  doi = {10.1103/PhysRevLett.104.216401},
  url = {https://link.aps.org/doi/10.1103/PhysRevLett.104.216401}
}

@article{Popescu2010,
 title = {Effective Band Structure of Random Alloys},
  author = {Popescu, Voicu and Zunger, Alex},
  journal = {Phys. Rev. Lett.},
  volume = {104},
  issue = {23},
  pages = {236403},
  numpages = {4},
  year = {2010},
  month = {Jun},
  publisher = {American Physical Society},
  doi = {10.1103/PhysRevLett.104.236403},
  url = {https://link.aps.org/doi/10.1103/PhysRevLett.104.236403}
}

@article{Allen2013,
  title = {Recovering hidden \ce{B}loch character: Unfolding electrons, phonons, and slabs},
  author = {Allen, P. B. and Berlijn, T. and Casavant, D. A. and Soler, J. M.},
  journal = {Phys. Rev. B},
  volume = {87},
  issue = {8},
  pages = {085322},
  numpages = {10},
  year = {2013},
  month = {Feb},
  publisher = {American Physical Society},
  doi = {10.1103/PhysRevB.87.085322},
  url = {https://link.aps.org/doi/10.1103/PhysRevB.87.085322}
}

@article{Kim2008,
title = {Origin of Anomalous Electronic Structures of \ce{E}pitaxial \ce{G}raphene on \ce{S}ilicon \ce{C}arbide},
  author = {Kim, Seungchul and Ihm, Jisoon and Choi, Hyoung Joon and Son, Young-Woo},
  journal = {Phys. Rev. Lett.},
  volume = {100},
  issue = {17},
  pages = {176802},
  numpages = {4},
  year = {2008},
  month = {Apr},
  publisher = {American Physical Society},
  doi = {10.1103/PhysRevLett.100.176802},
  url = {https://link.aps.org/doi/10.1103/PhysRevLett.100.176802}
}

@article{Medeiros2014,
   title = {Effects of extrinsic and intrinsic perturbations on the electronic structure of graphene: Retaining an effective primitive cell band structure by band unfolding},
  author = {Medeiros, Paulo V. C. and Stafstr\"om, Sven and Bj\"ork, Jonas},
  journal = {Phys. Rev. B},
  volume = {89},
  issue = {4},
  pages = {041407},
  numpages = {4},
  year = {2014},
  month = {Jan},
  publisher = {American Physical Society},
  doi = {10.1103/PhysRevB.89.041407},
  url = {https://link.aps.org/doi/10.1103/PhysRevB.89.041407}
}

@article{Maspero2017,
  author = {Maspero, R. and Sweeney, S. and Florescu, M.},
  title = {Unfolding the band structure of \ce{GaAsBi}},
  journal = {J. Phys.: Condens. Matter},
  volume = {29},
  pages = {075001},
  year = {2017},
  doi = {10.1088/1361-648X/aa50d7},
  url = {https://iopscience.iop.org/article/10.1088/1361-648X/aa50d7}
}

@article{Gajdos2006,
  title = {Linear optical properties in the projector-augmented wave methodology},
  author = {Gajdo\ifmmode \check{s}\else \v{s}\fi{}, M. and Hummer, K. and Kresse, G. and Furthm\"uller, J. and Bechstedt, F.},
  journal = {Phys. Rev. B},
  volume = {73},
  issue = {4},
  pages = {045112},
  numpages = {9},
  year = {2006},
  month = {Jan},
  publisher = {American Physical Society},
  doi = {10.1103/PhysRevB.73.045112},
  url = {https://link.aps.org/doi/10.1103/PhysRevB.73.045112}
}

@article{Hasan2010,
 title = {Colloquium: Topological insulators},
  author = {Hasan, M. Z. and Kane, C. L.},
  journal = {Rev. Mod. Phys.},
  volume = {82},
  issue = {4},
  pages = {3045--3067},
  numpages = {0},
  year = {2010},
  month = {Nov},
  publisher = {American Physical Society},
  doi = {10.1103/RevModPhys.82.3045},
  url = {https://link.aps.org/doi/10.1103/RevModPhys.82.3045}
}

@article{Sipe2000,
 title = {Second-order optical response in semiconductors},
  author = {Sipe, J. E. and Shkrebtii, A. I.},
  journal = {Phys. Rev. B},
  volume = {61},
  issue = {8},
  pages = {5337--5352},
  numpages = {0},
  year = {2000},
  month = {Feb},
  publisher = {American Physical Society},
  doi = {10.1103/PhysRevB.61.5337},
  url = {https://link.aps.org/doi/10.1103/PhysRevB.61.5337}
}

@article{Zak1989,
  title={Berry’s phase for energy bands in solids},
  author={Zak, Joshua},
  journal={Phys. Rev. Lett.},
  volume={62},
  number={23},
  pages={2747},
  year={1989},
  publisher={APS},
  doi = {https://doi.org/10.1103/PhysRevLett.62.2747},
  url = {https://journals.aps.org/prl/abstract/10.1103/PhysRevLett.62.2747}
}

@article{Xiao2010,
  title = {Berry phase effects on electronic properties},
  author = {Xiao, Di and Chang, Ming-Che and Niu, Qian},
  journal = {Rev. Mod. Phys.},
  volume = {82},
  issue = {3},
  pages = {1959--2007},
  numpages = {0},
  year = {2010},
  month = {Jul},
  publisher = {American Physical Society},
  doi = {10.1103/RevModPhys.82.1959},
  url = {https://link.aps.org/doi/10.1103/RevModPhys.82.1959}
}

@article{Novoselov2005,
  author = {Novoselov, K. S. and Geim, A. K. and Morozov, S. V. and Jiang, D. and Katsnelson, M. I. and Grigorieva, I. V. and Dubonos, S. V. and Firsov, A. A.},
  title = {Two-dimensional gas of massless \ce{D}irac fermions in graphene},
  journal = {Nature},
  volume = {438},
  pages = {197--200},
  year = {2005},
  doi ={https://doi.org/10.1038/nature04233},
  url = {https://www.nature.com/articles/nature04233}
}

@article{zhang2005,
  title={Experimental observation of the quantum \ce{H}all effect and \ce{B}erry's phase in graphene},
  author={Zhang, Yuanbo and Tan, Yan-Wen and Stormer, Horst L and Kim, Philip},
  journal={Nature},
  volume={438},
  number={7065},
  pages={201--204},
  year={2005},
  doi ={https://doi.org/10.1038/nature04235},
  url = {https://www.nature.com/articles/nature04235#citeas}, 
  publisher={Nature Publishing Group UK London}
}

@misc{SupplementalMaterial,
  note = {See Supplemental Material at \url{http://link.aps.org/supplemental/xxxx} for proof of $\left\| \hat{H}_{\mathrm{dis}}^{k',k} \right\| \ll \left\| \hat{H}_{\mathrm{dis}}^{k} \right\|$, detailed derivations of the spectral weight, and the Berry curvature calculation.}
}

@article{Li2009,
   title = {Topological \ce{A}nderson Insulator},
  author = {Li, Jian and Chu, Rui-Lin and Jain, J. K. and Shen, Shun-Qing},
  journal = {Phys. Rev. Lett.},
  volume = {102},
  issue = {13},
  pages = {136806},
  numpages = {4},
  year = {2009},
  month = {Apr},
  publisher = {American Physical Society},
  doi = {10.1103/PhysRevLett.102.136806},
  url = {https://link.aps.org/doi/10.1103/PhysRevLett.102.136806}
}

@article{Groth2009,
   title = {Theory of the Topological \ce{A}nderson Insulator},
  author = {Groth, C. W. and Wimmer, M. and Akhmerov, A. R. and Tworzyd\l{}o, J. and Beenakker, C. W. J.},
  journal = {Phys. Rev. Lett.},
  volume = {103},
  issue = {19},
  pages = {196805},
  numpages = {4},
  year = {2009},
  month = {Nov},
  publisher = {American Physical Society},
  doi = {10.1103/PhysRevLett.103.196805},
  url = {https://link.aps.org/doi/10.1103/PhysRevLett.103.196805}
}

@article{Chen2022,
  author = {Chen, W. and von Gersdorff, G.},
  title = {Measurement of interaction-dressed \ce{B}erry curvature and quantum metric in solids by optical absorption},
  journal = {SciPost Phys. Core},
  volume = {5},
  pages = {040},
  year = {2022},
  doi = {10.21468/SciPostPhysCore.5.3.040},
  url = {https://scipost.org/10.21468/SciPostPhysCore.5.3.040}
  
}

@article{Liu2024,
  title = {Effect of disorder on \ce{B}erry curvature and quantum metric in two-band gapped graphene},
  author = {Liu, Ze and Zhang, Zhi-Fan and Zhu, Zhen-Gang and Su, Gang},
  journal = {Phys. Rev. B},
  volume = {110},
  issue = {24},
  pages = {245419},
  numpages = {12},
  year = {2024},
  month = {Dec},
  publisher = {American Physical Society},
  doi = {10.1103/PhysRevB.110.245419},
  url = {https://link.aps.org/doi/10.1103/PhysRevB.110.245419}
}

@book{Mahan2013,
  author = {Mahan, G. D.},
  title = {Many-Particle Physics},
  publisher = {Springer},
  year = {2013}
}

@book{Sakurai2020,
  author = {Sakurai, J. J. and Napolitano, J.},
  title = {Modern Quantum Mechanics},
  publisher = {Cambridge University Press},
  year = {2020}
}

@article{Lordi2010,
title = {Charge carrier scattering by defects in semiconductors},
  author = {Lordi, Vincenzo and Erhart, Paul and \AA{}berg, Daniel},
  journal = {Phys. Rev. B},
  volume = {81},
  issue = {23},
  pages = {235204},
  numpages = {7},
  year = {2010},
  month = {Jun},
  publisher = {American Physical Society},
  doi = {10.1103/PhysRevB.81.235204},
  url = {https://link.aps.org/doi/10.1103/PhysRevB.81.235204}
}

@article{Kim2019,
    title = {Vertex corrections to the dc conductivity in anisotropic multiband systems},
  author = {Kim, Sunghoon and Woo, Seungchan and Min, Hongki},
  journal = {Phys. Rev. B},
  volume = {99},
  issue = {16},
  pages = {165107},
  numpages = {11},
  year = {2019},
  month = {Apr},
  publisher = {American Physical Society},
  doi = {10.1103/PhysRevB.99.165107},
  url = {https://link.aps.org/doi/10.1103/PhysRevB.99.165107}
}

@article{Kang2008,
  author = {Kang, J. and Bang, J. and Ryu, B. and Chang, K. J.},
  title = {Effect of atomic-scale defects on the low-energy electronic structure of graphene: Perturbation theory and local-density-functional calculations},
  journal = {Phys. Rev. B},
  volume = {77},
  pages = {115453},
  year = {2008}
}

@article{Resta1994,
title = {Macroscopic polarization in crystalline dielectrics: the geometric phase approach},
  author = {Resta, Raffaele},
  journal = {Rev. Mod. Phys.},
  volume = {66},
  issue = {3},
  pages = {899--915},
  numpages = {0},
  year = {1994},
  month = {Jul},
  publisher = {American Physical Society},
  doi = {10.1103/RevModPhys.66.899},
  url = {https://link.aps.org/doi/10.1103/RevModPhys.66.899}
}

@article{Resta2010,
  author = {Resta, R.},
  title = {Electrical polarization and orbital magnetization: The modern theories},
  journal = {J. Phys.: Condens. Matter},
  volume = {22},
  pages = {123201},
  year = {2010},
  doi = {10.1088/0953-8984/22/12/123201},
  url = {https://iopscience.iop.org/article/10.1088/0953-8984/22/12/123201}
}

@article{Tyner2024,
  author = {Tyner, A. C.},
  title = {\ce{B}erry\ce{E}asy: a \ce{GPU} enabled python package for diagnosis of nth-order and spin-resolved topology in the presence of fields and effects},
  journal = {J. Phys.: Condens. Matter},
  volume = {36},
  pages = {325902},
  year = {2024},
  doi = {10.1088/1361-648X/ad475f},
  url = {https://iopscience.iop.org/article/10.1088/1361-648X/ad475f}
}

@misc{PythTB2022,
  author = {Coh, S. and Vanderbilt, D.},
  title = {\ce{P}ython \ce{T}ight \ce{B}inding (\ce{P}yth\ce{T}\ce{B)}},
  howpublished = {Zenodo},
  year = {2022},
  doi = {10.5281/zenodo.12721316}
}

@incollection{Vanderbilt2006,
  author = {Vanderbilt, D. and Resta, R.},
  title = {Conceptual foundations of materials properties},
  booktitle = {Conceptual Foundations of Materials Properties},
  editor = {Louie, S. G. and Cohen, M. L.},
  publisher = {Elsevier},
  pages = {139--163},
  year = {2006}
}

@article{Xiao2007,
   title = {Valley-Contrasting Physics in Graphene: Magnetic Moment and Topological Transport},
  author = {Xiao, Di and Yao, Wang and Niu, Qian},
  journal = {Phys. Rev. Lett.},
  volume = {99},
  issue = {23},
  pages = {236809},
  numpages = {4},
  year = {2007},
  month = {Dec},
  publisher = {American Physical Society},
  doi = {10.1103/PhysRevLett.99.236809},
  url = {https://link.aps.org/doi/10.1103/PhysRevLett.99.236809}
}

@article{Yao2008,
  title = {Valley-dependent optoelectronics from inversion symmetry breaking},
  author = {Yao, Wang and Xiao, Di and Niu, Qian},
  journal = {Phys. Rev. B},
  volume = {77},
  issue = {23},
  pages = {235406},
  numpages = {7},
  year = {2008},
  month = {Jun},
  publisher = {American Physical Society},
  doi = {10.1103/PhysRevB.77.235406},
  url = {https://link.aps.org/doi/10.1103/PhysRevB.77.235406}
}

@article{Bianco2014,
  title = {How disorder affects the \ce{B}erry-phase anomalous \ce{H}all conductivity: A reciprocal-space analysis},
  author = {Bianco, Raffaello and Resta, Raffaele and Souza, Ivo},
  journal = {Phys. Rev. B},
  volume = {90},
  issue = {12},
  pages = {125153},
  numpages = {9},
  year = {2014},
  month = {Sep},
  publisher = {American Physical Society},
  doi = {10.1103/PhysRevB.90.125153},
  url = {https://link.aps.org/doi/10.1103/PhysRevB.90.125153}
}

@article{Olsen2015,
  title = {Valley \ce{H}all effect in disordered monolayer \ce{${\mathrm{MoS}}_{2}$} from first principles},
  author = {Olsen, Thomas and Souza, Ivo},
  journal = {Phys. Rev. B},
  volume = {92},
  issue = {12},
  pages = {125146},
  numpages = {12},
  year = {2015},
  month = {Sep},
  publisher = {American Physical Society},
  doi = {10.1103/PhysRevB.92.125146},
  url = {https://link.aps.org/doi/10.1103/PhysRevB.92.125146}
}

@article{Martiny2019,
title = {Tunable valley \ce{H}all effect in gate-defined graphene superlattices},
  author = {Martiny, Johannes H. J. and Kaasbjerg, Kristen and Jauho, Antti-Pekka},
  journal = {Phys. Rev. B},
  volume = {100},
  issue = {15},
  pages = {155414},
  numpages = {11},
  year = {2019},
  month = {Oct},
  publisher = {American Physical Society},
  doi = {10.1103/PhysRevB.100.155414},
  url = {https://link.aps.org/doi/10.1103/PhysRevB.100.155414}
}

\end{document}